\documentstyle[preprint,aps,epsfig]{revtex}
\begin{document}
\preprint{FSU-SCRI-97-96}
\title{Pion Cloud Contributions to the Proton's Weak Magnetic Form
Factor}
\author{Simon Capstick and D. Robson}
\address{Supercomputer Computations Research Institute and Department
of Physics\\
Florida State University, Tallahassee, FL 32306}
\date{\today}
\maketitle
\begin{abstract}
We report on the contributions to the nucleon weak magnetic and
electric form factors arising from isospin-breaking form factors. The
relativized quark model used in the calculations includes isospin
impurities in the large basis nucleon wavefunctions and uses
constituent quark currents with both Pauli and Dirac
components. Significant contributions are found to arise from the
Pauli component when it is interpreted as arising from pion-baryon
loops. Including these pion cloud contributions reduces the strange
magnetic form factor extracted from the recent SAMPLE measurement of
the proton's weak neutral magnetic form factor at low $Q^2$ by 35 to
52\%.
\end{abstract}

There has been considerable interest for several years now in the
possibility of measuring the effects of the strange quark-antiquark
($\overline{s}s$) pair in the nucleon using the parity-violating
asymmetry in elastic electron scattering from the
nucleon~\cite{expts}. This year the first measurement of this
asymmetry was reported on by the SAMPLE Collaboration~\cite{SAMPLE}
and clearly demonstrates the feasibility of seeing asymmetries of a
few parts per million. The experiment was performed at the MIT/Bates
Linear Accelerator Center using 200 MeV incident electrons elastically
scattered from the proton at backward angles with an average $Q^2$ of
0.1 (GeV/c)$^2$. With this choice of kinematics the neutral weak form
factor for the proton $G^Z_M$ is enhanced relative to the
corresponding electric term and can be extracted from the asymmetry
using the relation
\begin{equation}
A={\sigma_R-\sigma_L\over \sigma_R+\sigma_L}
=-{G_F Q^2\over \pi\alpha\sqrt{2}}
\times\left\{ {\epsilon G_E^\gamma G_E^Z + \tau G_M^\gamma G_M^Z -
\frac{1}{2}(1-4{\rm sin}^2\theta_W)\epsilon^\prime G_M^\gamma G_A^Z
\over \epsilon (G_E^\gamma)^2+\tau (G_M^\gamma)^2}
\right\},
\label{A}
\end{equation}
where $\epsilon$, $\tau$, and
$\epsilon^\prime=\sqrt{\tau(1+\tau)(1-\epsilon^2)}$ are kinematic
quantities and $Q^2>0$ is the four momentum transfer squared. The
electromagnetic electric and magnetic form factors for the proton are
denoted by $G_E^\gamma$ and $G_M^\gamma$ respectively. The
corresponding electric and magnetic neutral weak form factors are
denoted by the superscript $Z$. The numerator also
contains the neutral weak axial-vector form factor $G_A^Z$ which
causes a significant correction in the extracted value of $G_M^Z$.

The interest in the neutral weak magnetic form factor lies in the
simple relationship it has to the strange magnetic form factor $G^S_M$
when certain simplifying assumptions are made. To tree level order,
and assuming no isospin breaking in the nucleon, one finds using the
notation of Ref.~\cite{SAMPLE}
\begin{equation}
G_M^Z=\frac{1}{4}(G^p_M-G^n_M)-{\rm sin}^2\theta_W G^p_M-\frac{1}{4}G^S_M,
\label{GMZ}
\end{equation}
where $\theta_W$ is the weak mixing angle known from a recent high
precision measurement~\cite{PDG}: ${\rm sin}^2\theta_W(M_Z)=0.2315\pm
0.0004$. Electroweak radiative corrections have been applied to
Eq.~(\ref{GMZ}) in the report on the SAMPLE experiment. However, the
effects of isospin breaking are not considered in their analysis.

In this Letter we report on a detailed calculation of the additional
form factors arising from isospin breaking and follow the notation of
Ref.~\cite{DPoll}. The isoscalar ($u+d$) and isovector ($u-d$) Dirac
($F_1$) and Pauli ($F_2$) form factors for the proton and neutron are
defined by
\begin{eqnarray}
\langle N(p^\prime)\vert &&\frac{1}{2}\left(\overline{u}\gamma_\mu u
-\overline{d}\gamma_\mu d\right)\vert N(p)\rangle\nonumber\\
&&\equiv \overline{\cal U}(p^\prime)
\left[\frac{1}{2}\left(^{u-d}F_1^{p+n}\pm ^{u-d}F_1^{p-n}\right)
\gamma_\mu
+\frac{1}{2}\left(^{u-d}F_2^{p+n}\pm ^{u-d}F_2^{p-n}\right)
i\sigma_{\mu\nu}{q^\nu\over 2M_N}\right]{\cal U}(p)\nonumber\\
\langle N(p^\prime)\vert &&\frac{1}{6}\left(\overline{u}\gamma_\mu u
+\overline{d}\gamma_\mu d\right)\vert N(p)\rangle\nonumber\\
&&\equiv\overline{\cal U}(p^\prime)
\left[\frac{1}{2}\left(^{u+d}F_1^{p+n}\pm ^{u+d}F_1^{p-n}\right)
\gamma_\mu
+\frac{1}{2}\left(^{u+d}F_2^{p+n}\pm ^{u+d}F_2^{p-n}\right)
i\sigma_{\mu\nu}{q^\nu\over 2M_N}\right]{\cal U}(p),
\label{Nisosv}
\end{eqnarray}
where the sign is $+$ for the proton and $-$ for the neutron. The
corresponding Sachs form factors are given by the linear combinations
$G_E=F_1-\tau F_2$ and $G_M=F_1+F_2$. For the neutral weak magnetic
form factor of the proton Eq.~(\ref{GMZ}) is modified by replacing the
strange magnetic form factor $G^S_M$ by the linear combination
$G^S_M+ ^{u+d}G_M^{p-n}- ^{u-d}G_M^{p+n}$. Note that the strange form
factor arises from the current operator $\overline{s}\gamma_\mu s$
whereas the isospin-breaking terms arise from isovector and isoscalar
currents with factors of $\frac{1}{2}$ and $\frac{1}{6}$ respectively.

The calculations of the form factors $^{u+d}G_{E,M}^{p-n}$ and
$^{u-d}G_{E,M}^{p+n}$ utilize the relativized quark model based on the
light-cone formalism of Ref.~\cite{CK} which produces realistic mixed
wavefunctions for each nucleon by solving the three-body problem in a
large basis of harmonic oscillator states. The model allows for the
constituent quarks to have (a) different masses for the $u$ and $d$
quarks, (b) electromagnetic interactions (Coulomb + hyperfine) between
the quarks, and (c) both Dirac and Pauli electromagnetic form factors
corresponding to the substitution
\begin{equation}
\overline{q}e_q\gamma_\mu q \to \overline{q}_c e_q\gamma_\mu f_{1q}
q_c + \overline{q}_c \kappa^{(N)}_q i\sigma_{\mu\nu}{q^\nu\over 2m_q}
f_{2q} q_c,
\label{qcurrent}
\end{equation}
where $q_c$ denotes a constituent quark appropriate to the model. In
the $u_c$, $d_c$ sector the Dirac component can be written as
isoscalar and isovector currents for each nucleon as in
Eqs.~(\ref{Nisosv}). The Pauli component term involves anomalous
moments $\kappa^{(N)}_q$ which depend on the quark flavor {\it and}
the nucleon flavor. This is an important assumption which is
understandable if one assumes that the Pauli term arises from
interactions with the charged pion cloud corresponding to the
$\overline{u}u$ and $\overline{d}d$ loops in Fig.~\ref{loops}.

In these calculations there are isospin-breaking effects associated
with both of the quark form factors. The Dirac quark component has
isospin violations because of the charge dependence of the Hamiltonian
used. In the expansion of the proton and neutron wavefunctions into a
basis of states with good isospin given by
\begin{equation}
\Psi^{J^P=\frac{1}{2}^+}_N=\sum_\alpha
a_\alpha(N)\phi^{\frac{1}{2}^+}_\alpha(T=\frac{1}{2},T_z)
+\sum_\beta
b_\beta(N)\phi^{\frac{1}{2}^+}_\beta(T=\frac{3}{2},T_z),
\end{equation}
there are mirror violations from the fact that $a_\alpha(p)$ is
slightly different from $a_\alpha(n)$ as well as the usual violations
{\it via} the admixture of the $T=\frac{3}{2}$ delta
configurations. The former $T=\frac{1}{2}$ `dynamic distortion' is
larger than the $T=\frac{3}{2}$ violation and both contribute to the
isospin-breaking form factors calculated here. The rms averages when
95 basis states (corresponding to oscillator quanta up to
$8\hbar\omega$) are used are $A=(\sum_\alpha 
[a_\alpha(p)-a_\alpha(n)]^2)^{\frac{1}{2}}=1.57\times 10^{-3}$ and 
$B=(\sum_\beta 
[b_\beta(p)+b_\beta(n)]^2)^{\frac{1}{2}}=0.88\times 10^{-3}$, yielding
a total average violation $C=(A^2+B^2)^{\frac{1}{2}}=1.8\times
10^{-3}$. The isospin-breaking form factors $^{u+d}G^{p-n}$ and 
$^{u-d}G^{p+n}$ in our calculations do show contributions from the
Dirac quark component which are of the order of magnitude of $C$.

Violations arising from the Pauli quark component are of considerable
interest because they are entirely analogous to one of the mechanisms
proposed~\cite{Musolf} for the strange form factors of the nucleon. In
the case of $\overline{s}s$ loops the $\overline{s}$ coalesces with a
$u$ quark to form a $K^+$ and the $s$ joins with the spectator pair of
quarks to form a $\Lambda$ or $\Sigma$ baryon. The difference in the
momentum distributions of an $\overline{s}$ in the $K^+$ and an $s$ in
the hyperons causes the strange form factors to be non-zero. Such
non-perturbative effects from the loops involving the lightest meson
and baryons are expected at low $Q^2$. By analogy we see in
Fig.~\ref{loops} that the pion-nucleon and pion-delta loops should
have strong differences in the momentum distributions for the
$\overline{u}$, $\overline{d}$ in the pion and their $u$, $d$ partners
in the nucleon or delta. In our parameterization of the anomalous
moments we assume the photon or $Z$ interacts predominantly with the
pion. This is expected to be a valid assumption because for the light
baryons the pion is in a $p$ orbit and with its small mass it will
dominate the quark Pauli component contributions to both the nucleonic
charge radius and the nucleonic anomalous magnetic moment. We also
require the relations $\kappa^{(p)}_u=-\kappa^{(n)}_d$ and
$\kappa^{(p)}_d=-\kappa^{(n)}_u$ due to charge symmetry which relates
$\pi^+$ to $\pi^-$ and $u$ to $d$ when transforming the proton into a
neutron by the charge symmetry operator, {\it i.e.} a 180$^o$ rotation
about the $y$ axis in isospin space. These relations can be visualized
by noting the similar hadron content of the {\it left} diagrams in
Fig.~\ref{loops}, which in our model give $\kappa^{(p)}_u$ and
$\kappa^{(n)}_d$, and noting the sign of the photon or $Z$ coupling to
the pion in the loop, and similarly for the {\it right} diagrams. Note
that the right diagrams will be suppressed due to the lack of
a nucleon--pion intermediate state. Previous calculations using Pauli
quark components~\cite{Romans} appear to have ignored the importance
of charge symmetry which relates constituent $u$ quarks in the proton
to constituent $d$ quarks in the neutron and {\it vice-versa}.

The parameters for the quark-model Hamiltonian are determined by
fitting the baryon spectrum, and the quark mass difference
$m_d-m_u=3.6$ MeV is fixed by fitting the nucleon mass difference
$M_n-M_p=1.29$ MeV. The parameters of the current quark operators
$\kappa^{(p)}_u$, $\kappa^{(p)}_d$, and the form factors
$f_i(Q^2)=(1+Q^2/\Lambda_i^2)^{-i}$ with $i=1,2$ are determined by
fitting the observed proton and neutron electric and magnetic form
factors. These fits deviate from experiment by amounts which are
typically of the order of 5\% for $Q^2$ up to 1 (GeV/c)$^2$, except
for the neutron form factors which are not known to this
accuracy. The values of $\kappa^{(p)}_u$, $\kappa^{(p)}_d$, and the
$\Lambda_i$ are predominantly determined by the nucleon magnetic
moments and the neutron rms charge radius. Without the pion cloud
contributions contained in the Pauli quark current the magnitude of
the nucleon magnetic moments are underpredicted by about 0.6
n.m. Similarly we find that the pion cloud provides about 50\% of the
neutron charge radius. Using the parameters
$\kappa^{(p)}_u=0.12=-\kappa^{(n)}_d$,
$\kappa^{(p)}_d=-0.04=-\kappa^{(n)}_u$, and $\Lambda_1^2=1.22$
GeV$^2$, $\Lambda_2^2=0.30$ GeV$^2$, we obtain the $u+d$
isospin-breaking form factors shown in Figure~\ref{u+dGp-n}.

We have not plotted the corresponding $u-d$ terms in
Fig.~\ref{u+dGp-n} since, as shown in Table~\ref{fftable}, their
numerical values are too small to be observed. The numerical results
in Table~\ref{fftable} also include the form factor calculations for
the Dirac current alone (all $\kappa^{(N)}_q=0$) and show that the
large values of the $u+d$ terms arise almost entirely from the
pion-cloud contributions to the Pauli term. The isovector nature of
the pion cloud arises because we have assumed that the contribution
from the quark partner in the $\overline{q}q$ loops of
Fig.~\ref{loops} are suppressed relative to the antiquark partner
residing in the pion. If one assumed that this quark contribution was
the same as that of its antiquark partner then the anomalous quark
moments would be proportional to the constituent quark charges and the
same for each flavor in the neutron and proton, {\it i.e.} the Pauli
term would be like the Dirac term and yield only small $u+d$
isospin-breaking form factors (of the order of $C=2\times 10^{-3}$)
for the nucleon. We contend that the dynamical symmetry breaking
mechanism which places the antiquark in a Goldstone boson (small mass)
and its quark partner in a nucleon or delta baryon will produce Pauli
quark form factors which have anomalous moments of the type used
here. The ratio of the $u$ and $d$-quark anomalous moments arising
from the diagrams in Fig.~\ref{loops} have been estimated by us to be
approximately $-2$ to $-3$. Calculations with a ratio of $-2$ reduce
the magnitude of the form factors in Fig.~\ref{u+dGp-n} by about
30\%. At $Q^2=0.1$ (GeV/c)$^2$ we obtain a contribution to the weak
neutral form factor from the pion cloud of between -0.02 and -0.03
n.m. Inclusion of this contribution in the SAMPLE collaboration
results would reduce the extracted value of the strange magnetic form
factor at this $Q^2$ ({\it i.e.} $G^S_M=+0.23\pm 0.09 \pm 0.04 \pm
0.05$) by 35 to 52\%.

In summary we believe the pion cloud effects calculated
here are comparable in size to the kaon cloud effects expected for the
corresponding strange form factors for the nucleon. We expect the
$Q^2$ dependences of the pion and kaon cloud contributions to be
different due to the significant difference in their masses. It will
be interesting to see if the planned experiments~\cite{expts} can
provide information on both the pion and kaon cloud components of the
nucleon and $^4$He. Considerations of other sources of isospin
breaking and strange form factors~\cite{JaffeSpeth}, such as photon
coupling to the vector mesons $\rho$, $\omega$, $\phi$, {\it etc.},
will be needed. In particular the effects of both $\rho-\omega$ mixing
on $^{u\pm d}G_{E,M}^{p\mp n}$ and of $\omega-\phi$ mixing on
$G^S_{E,M}$ need to be better understood. A more detailed account of
this work including applications to $^4$He will be reported elsewhere.

The authors are grateful to Prof.~B.D.~Keister for his essential
contribution to the calculation of the light-cone electromagnetic form
factors. This work was supported in part by the U.S. Department of
Energy under Contract No.~DE-FG05-86ER40273 (SC, DR), and the Florida
State University Supercomputer Computations Research Institute which
is partially funded by the Department of Energy through Contract
DE-FC05-85ER250000 (SC).

\begin{table}
\caption{Numerical values of the isospin-breaking form factors
$^{u+d}G_{E,M}^{p-n}$ and $^{u-d}G_{E,M}^{p+n}$ for $0.1\leq Q^2\leq
0.8$ GeV$^2$ for (a) all $\kappa^{(N)}_q=0$ and (b)
$\kappa^{(p)}_u=0.12=-\kappa^{(n)}_d$, and
$\kappa^{(p)}_d=-0.04=-\kappa^{(n)}_u$ as described in the text. Units
of $G_E$ are $10^{-3}$, and units of $G_M$ are $10^{-3}$ n.m.}
\label{fftable}
\begin{tabular}{c|rcrr|rcrr}
&&\multicolumn{2}{c}{(a)}&&&\multicolumn{2}{c}{(b)}\\ 

$Q^2$ (GeV/c)$^2$ & $^{u+d}G_E^{p-n}$ & $^{u+d}G_M^{p-n}$ &
$^{u-d}G_E^{p+n}$ & $^{u-d}G_M^{p+n}\phantom{\ }$ & $^{u+d}G_E^{p-n}$ &
$^{u+d}G_M^{p-n}$ & $^{u-d}G_E^{p+n}$ & $^{u-d}G_M^{p+n}$\\[6pt]
\tableline
&&&&&\\[-6pt]
0.0 &    0.0 & 3.0 &    0.0 & 1.3$\phantom{\ }$  &  0     & 250 & 0  & 3.7\\
0.1 &    0.0 & 2.3 & $-1.5$ & 2.5$\phantom{\ }$  & $-27$ & 124 & -2.1 & 3.5\\
0.2 & $-0.1$ & 1.6 & $-2.2$ & 2.0$\phantom{\ }$  & $-30$ & 56 & -2.8 & 2.3\\
0.3 & $-0.2$ & 1.2 & $-2.6$ & 1.4$\phantom{\ }$  & $-28$ & 31 & -3.2 & 1.5\\
0.4 & $-0.2$ & 0.8 & $-2.7$ & 0.9$\phantom{\ }$  & $-25$ & 18 & -3.2 & 0.9\\
0.5 & $-0.2$ & 0.6 & $-2.8$ & 0.5$\phantom{\ }$  & $-23$ & 11 & -3.2 & 0.4\\
0.6 & $-0.2$ & 0.4 & $-2.7$ & 0.2$\phantom{\ }$  & $-20$ & 6 & -3.0 & 0.1\\
0.7 & $-0.2$ & 0.3 & $-2.5$ & $-0.1\phantom{\ }$ & $-18$ & 3 & -2.9 & $-0.1$\\
0.8 & $-0.2$ & 0.2 & $-2.4$ & $-0.2\phantom{\ }$ & $-15$ & 2 & -2.7 & $-0.3$\\
&&&&&\\[-6pt]
\end{tabular}
\end{table}

\vbox{          
\begin{figure}
\epsfig{file=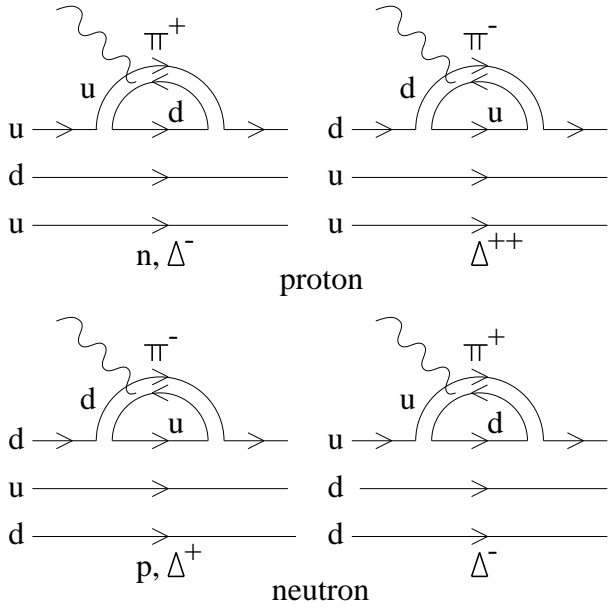,width=8cm,angle=-90}
\vspace*{0.5cm}
\caption{Virtual $\overline{q}q$ pairs which contribute to the Pauli
component for constituent quarks in each nucleon.}
\label{loops}
\end{figure}
} 

\vbox{          
\begin{figure}
\epsfig{file=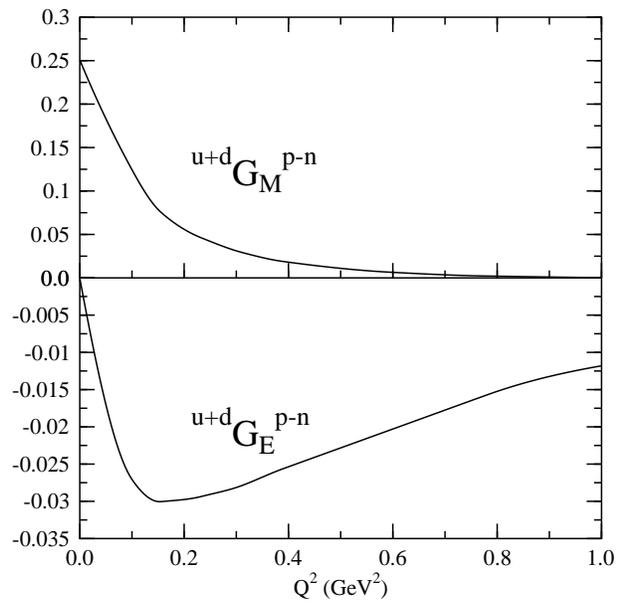,width=10.0cm,angle=0}
\vspace*{-0.25cm}
\caption{Predictions for the isospin-breaking form factors
$^{u+d}G_E^{p-n}$, $^{u+d}G_M^{p-n}$ which contribute to the neutral
weak electric and magnetic proton form factors respectively. Note the
different vertical scales above and below the $Q^2$ axis.}
\label{u+dGp-n}
\end{figure}
}       

\end{document}